\documentclass{ws-ijmpa}
\usepackage[super,compress]{cite}
\usepackage{graphicx}
\usepackage{multirow}
\usepackage{bm}
\usepackage{amsmath}
\usepackage{amssymb}
\usepackage{mathrsfs}
\usepackage{subfigure}
\usepackage{epstopdf}
\begin{document}
\markboth{T. Inagaki, D. Kimura, H. Kohyama}
{Next to leading order calculation with dimensional regularization 
in NJL model}

%
\catchline{}{}{}{}{}
%

\title{ Next to leading order calculation with dimensional regularization 
in Nambu--Jona-Lasinio Model
}

\author{T. Inagaki$^a$, D. Kimura$^b$, H. Kohyama$^c$}

\address{$^a$Information Media Center, Hiroshima University,
Higashi-Hiroshima, Hiroshima
739-8521, Japan}
\address{$^b$General Education, Ube National College of Technology,
Ube, Yamaguchi 755-8555, Japan}
\address{$^c$
Department of Physics, Kyungpook National University,
Daegu 702-701, Korea
}

\maketitle


\begin{abstract}
The Nambu--Jona-Lasinio model is investigated in the $1/N_c$ expansion 
with the dimensional regularization. At the four-dimensional limit
the meson propagators have simple forms in the leading order of the
$1/N_c$ expansion. Thus the next to leading order calculation reduces to
an ordinary one loop calculation. Here we obtain an explicit form 
of the $1/N_c$ correction and numerically evaluate the $N_c$ dependence 
for the gap equation.

\keywords{low energy effective theory; dimensional regularization; 
$1/N_c$ expansion.}
\end{abstract}

\ccode{PACS numbers:11.30.Qc, 12.39.-x}


\section{Introduction}
The strong interaction between quarks and gluons are described by
quantum chromodynamics (QCD). Because of the asymptotic freedom, the
non-perturbative effect is essential for low energy phenomena in QCD.
Nambu and Jona-Lasinio introduce a four-fermion interaction to study
the meson properties with the strong interaction \cite{NJL}. The NJL
model has a similar symmetry behavior to QCD. It is used as one of
the effective models of QCD for low energy \cite{Vogl:1991qt,
Klevansky:1992qe, Hatsuda:1994pi}.

The critical behavior for quarks and gluons is often discussed at the
leading order of $1/N_c$ expansion in the NJL model. The number of colors,
$N_c$, is three in the real world. Thus the quantitative properties
can be determined with an accuracy of about $30\%$ in the
leading order of $1/N_c$ expansion. There is another ambiguity for the
results in the NJL model. Since the four-fermion interaction is
irrelevant, it is necessary to regularize the model in four dimensions.
The result depends on the regularization parameter \cite{Fujihara:2008ae,
Inagaki:2010nb, Inagaki:2011uj, Inagaki:2012re}. 
The parameter is usually fixed to be consistent with the light meson properties.

In our previous study Ref. [\citen{Inagaki:2013hya}], we considered the NJL
model with the dimensional regularization in the four-dimensional limit.  
It is found that the calculations of meson mass and decay constant are
simplified in the leading order of the $1/N_c$ expansion.
Then it is expected that the analysis may also be simple if we
proceed to the next to leading order level.  

In this letter, we discuss the next to leading order of the $1/N_c$
expansion, and study the possible $N_c$ dependence in the model predictions.
This is interesting because the $N_c$ dependence is absorbed in the other
parameters in the leading order, and it first appears from the next to
leading order.

\section{$1/N_c$ expansion in the NJL model}
\label{njl_model}
In this section we briefly review the $1/N_c$ expansion in 
the NJL model for up and down quarks.
The Lagrangian of two-flavor NJL model is given as,
\begin{eqnarray}
 \mathcal{L} = 
 \sum_{j=1}^{N_c}\bar{\psi}_j\left( i \partial\!\!\!/ - m \right) \psi_j
  + g\sum_{j=1}^{N_c}\left[ (\bar{\psi}_j\psi_j)^2
  + (\bar{\psi}_j i\gamma_5 \tau^a \psi_j)^2 \right] ,
\label{L_NJL}
\end{eqnarray}
where $m$ is the current quark mass, $\tau^a$ is the Pauli matrices in
the flavor space, and $N_c$ is the number of colors which is treated
as one of model parameters in this paper. We neglect the mass difference
between up and down quarks. In the scheme of the $1/N_c$ expansion we 
take the large $N_c$ limit as $N_c g$ fixed.

By applying the auxiliary field method, we can evaluate the
generating functional from Eq. (\ref{L_NJL}) as
\begin{eqnarray}
Z &=& \int {\cal D}\bar{\psi} {\cal D}\psi {\cal D}\sigma {\cal D}\pi
 \exp \left[i\int d^4x \biggl\{ \bar{\psi}
  \left( i \partial\!\!\!/ - m \right) \psi
  - \bar{\psi}\left( \sigma + i\gamma_5 \tau^a \pi^a \right) \psi 
\right.  \nonumber \\
  &&\left. \left.- \frac{1}{4g} \left[ \sigma^2 + (\pi^a)^2 \right]
  \right\} \right] 
\\
&=& \int {\cal D}\sigma {\cal D}\pi
 \exp \left[i I(\sigma,\pi) \right] ,
\label{Z}
\end{eqnarray}
with
\begin{equation}
I(\sigma,\pi) = -\frac{1}{4g} \int d^4x 
\left\{ \sigma^2 + (\pi^a)^2 \right\}
 - i \ln {\rm Det} \left( i\partial\!\!\!/ - m - \sigma 
  - i\gamma_5 \tau^a \pi^a   \right) .
\label{I}
\end{equation}
In Eq. (\ref{I}) ``Det'' takes the color, flavor, spinor and space-time
indices. $I(\sigma, \pi)$ can formally be expanded around
the classical solution $\sigma_0$ and $\pi_0$,
\begin{eqnarray}
I(\sigma, \pi) &=& I(\sigma_0, \pi_0) 
 + \frac12 \frac{\delta^2 I(\sigma_0, \pi_0)}
 {\delta \sigma^2} (\sigma - \sigma_0)^2  \nonumber \\
&& + \frac12 \frac{\delta^2 I(\sigma_0, \pi_0)}
 {\delta \pi^2} (\pi - \pi_0)^2 + \cdots 
\label{I_exp}
\end{eqnarray}

In the leading order of the expansion, the effective potential,
$V=(i/\int d^4 x) \ln Z$, is written by
\begin{eqnarray}
V_0(\sigma, \pi) = \frac{1}{4g}\left\{ \sigma^2 + (\pi^a)^2 \right\}
- \int \frac{d^4 k}{i(2\pi)^4}
{\rm tr}\ln ( k\!\!\!/ -m -\sigma -i\gamma_5 \tau^a \pi^a ) ,
\label{V_0}
\end{eqnarray}
where ``{\rm tr}'' takes the color, flavor and spinor indices.

From the stable condition $\partial V/\partial \sigma=0$
at $\sigma=\sigma_0$, one can
derive the gap equation whose leading order form becomes 
\begin{equation}
  \sigma_0 =
  2N_f g\ i {\rm tr} S(m^*),
\label{gap} 
\end{equation}
where $m^* = m + \sigma_0$ and
\begin{eqnarray}
 i\, {\rm tr} S(m^*) &=& -\int \frac{d^4k}{i(2\pi)^4} 
 {\rm tr} \frac{1}{k\!\!\!/-m_i^*+i \varepsilon}.
\label{trS} 
\end{eqnarray}
``tr'' in the integral 
denotes the trace with respect to the spinor
and color indices.

In the leading order of the $1/N_c$ expansion, the propagator of pion
is given by \cite{Inagaki:2007dq}
\begin{eqnarray}
\Delta_\pi(p^2) &=& \frac{2g}{1-4g\Pi_5(p^2)}.
\label{pro_pi}
\end{eqnarray}
The pion mass is 
determined at the pole position of the propagator, namely,
\begin{equation}
  1 - 4 g \Pi_5(p^2=m_\pi^2) = 0 ,
\label{m_pi} 
\end{equation}
where
\begin{eqnarray}
\Pi_5(p^2) &=& -\int \frac{d^4k}{i(2\pi)^4} 
 {\rm tr}\left[i\gamma_5 S(k) i\gamma_5 S(k-p) \right] \nonumber \\
 &=& \frac{i{\rm tr}S}{m^*} + \frac12 p^2 J(p^2) ,
\label{Pi_5}
\end{eqnarray}
with
\begin{eqnarray}
J(p^2) &=& \int \frac{d^4k}{i(2\pi)^4} 
 {\rm tr} \frac1{(k^2-m^{*2}) \left\{ (k-p)^2 - m^{*2} \right\}}.
\end{eqnarray}
Similarly, the propagator of sigma meson is given as \cite{Inagaki:2007dq}
\begin{eqnarray}
\Delta_\sigma(p^2) &=& \frac{2g}{1-4g\Pi_s(p^2)} ,
\label{pro_sig}
\end{eqnarray}
where
\begin{eqnarray}
\Pi_{\mathrm s}(p^2) &=& -\int \frac{d^4k}{i(2\pi)^4} 
 {\rm tr}\left[S(k) S(k-p) \right].
\end{eqnarray}
The mass of the sigma meson is evaluated at the pole position as well.

\section{Parameter fixing in the $4$ dimensional limit}
\label{parameter}
Since the model is not renormalizable, the predictions depend on the
regularization procedure. Here we shall employ the dimensional
regularization in the $4$ dimensional limit \cite{Inagaki:2013hya}.
In this method, although the intermediate integrals diverge, the model
predictions turn out to be finite thanks to the parameter fixing.

In our model treatment, we have three parameters, $m, g$ and $M$.
$m$ is the current quark mass, $g$ is the four point coupling and $M$
is the mass rescaling parameter.
In this paper we take $m$ as the input parameter, the remaining two
parameters will be fixed by using the physical observables. We try to fix
these parameters with $m_\pi$ and $ f_\pi$
in the leading order of the $1/N_c$ expansion.

In the $4$ dimensional limit, $D(\equiv 4-2\epsilon) \to 4$, the
pion propagator can be written by \cite{Inagaki:2007dq}
\begin{eqnarray}
\Delta_\pi(p^2) 
= -\frac{Z_\pi M^{2\epsilon}}{p^2- m_\pi^2} .
\label{pro_pi_4d}
\end{eqnarray}
where the wave function renormalization is
\begin{eqnarray}
Z_\pi^{-1} = \frac{N_c}{4\pi^2 \epsilon} M^{2\epsilon} .
\end{eqnarray}
Then the pion propagator becomes
\begin{equation}
\Delta_\pi(p^2) = -\frac{4\pi^2\epsilon}{N_c} \frac{1}{p^2-m_\pi^2}.
\label{pro_pi2}
\end{equation}
The pion decay constant is also calculated in the same procedure,
\begin{eqnarray}
f_\pi^2 = \frac{N_c}{4\pi^2 \epsilon} M^{2\epsilon} \sigma_0^2.
\label{f_pi}
\end{eqnarray}

Next, let us consider the gap equation.  Performing the integral
Eq. (\ref{trS}) in the $4$ dimensional limit and inserting the
result into Eq. (\ref{gap}), we obtain
\begin{equation}
  \sigma_0 =
  -\frac{N_f N_c g}{2\pi^2 \epsilon} m^{*3}.
\label{gap_4d} 
\end{equation}
Since the following relation is derived from Eq. (\ref{Pi_5}) 
\begin{eqnarray}
 \Pi_5(m_\pi^2) 
 =
 \frac{N_c}{8\pi^2 \epsilon} (m_\pi^2 - 2m^{*2}) ,
\end{eqnarray}
we get, with the help of Eqs.  (\ref{m_pi}) and (\ref{gap_4d}),
\begin{equation}
\sigma_0 = -m - \frac{m_\pi^2}{4m} 
 \left\{ 1 + \sqrt{1+\frac{8m^2}{m_\pi^2}}
 \right\} ,
\label{sig}
\end{equation}
and
\begin{equation}
g = - \frac{\pi^2 \epsilon}{N_c} \frac{\sigma_0}{m^{*3} } .
\label{g}
\end{equation}
Equation (\ref{f_pi}) reads the relation
\begin{equation}
\lim_{\epsilon\to 0} M^{2\epsilon}
= \lim_{\epsilon\to 0} \frac{4\pi^2 \epsilon}{N_c}
 \frac{f_\pi^2}{\sigma_0^2} .
\label{M0}
\end{equation}
It is interesting to note that the parameter $M$ approaches to
$0$ in the $4$ dimensional limit. This is the essentially important
property of the mass rescaling parameter; we can control the divergent
integrals by virtue of the adjustment of mass dimensions through $M$
\cite{Inagaki:2013hya}.

Using Eqs.  (\ref{gap}), (\ref{sig}) and (\ref{M0}), we obtain the 
chiral condensate,
\begin{eqnarray}
  \langle \bar{u}u \rangle_0
  &=& -M^{2\epsilon} (i {\rm tr} S) \nonumber \\
  &=& \frac{f_\pi^2 m_\pi^4}{8m^3}
   \left\{ 1 - \sqrt{1 + \frac{8m^2}{m_\pi^2}}
   \right\} + O(1/N_c) .
\label{uu}
\end{eqnarray}
Note that the expansion of Eq. (\ref{uu}) in powers of $m$ leads the
Gell-Mann--Oakes--Renner relation\cite{GellMann:1968rz},
$\langle \bar{u} u \rangle_0 \simeq - f_\pi^2 m_\pi^2 /(2m)$ .

In the similar manner, we have for the sigma meson propagator
\begin{eqnarray}
\Delta_\sigma(p^2) = -\frac{4\pi^2\epsilon}{N_c} \frac{1}{p^2-m_{\sigma 0}^2} , 
\label{pro_sig2}
\end{eqnarray}
where the sigma mass can be written 
\begin{equation}
m_{\sigma 0}^2 = \frac{m_\pi^2}{2m^2} \left(
 m_\pi^2 + 6m^2 + m_\pi \sqrt{m_\pi^2 + 8m^2} \right)
\label{m_sig} .
\end{equation}
For $m_\pi=135$MeV and $m=5.0$MeV, we get
$m_{\sigma 0} \simeq 3700$MeV. This value corresponds with the one
obtained in Ref. [\citen{Inagaki:2007dq}], where the realistic value, $400-550$MeV \cite{Beringer:1900zz}, is found in different dimension
around $D\simeq 2$.
Thus, in the four dimensional limit with the leading order calculation
we find the deviation on the value of $m_\sigma$. This deviation may
be modified by considering the next to leading order of $1/N_c$ expansion,
which will be discussed in the following.

\section{Next to leading order of the $1/N_c$ expansion}
\label{next}
We have presented the analysis in the $4$ dimensional limit within
the leading order of $1/N_c$ expansion. We shall carry on the
calculations up to the next to leading order in this section.

\subsection{Formula for the next to leading order}
\label{formula}
We consider next to leading order of the $1/N_c$ expansion for
the effective potential by using the technique of auxiliary field
method Ref. [\citen{Kashiwa:2003rj}].  The second and third term
of Eq. (\ref{I_exp}) are
\begin{eqnarray}
\frac{\delta^2 I}{\delta \sigma(x) \delta \sigma(y)} 
\bigg|_{\sigma=\sigma_0,\pi=\pi_0} 
= -\frac{1}{2g}\delta^4(x-y) + i\, {\rm tr} [S(x,y;m^*)S(y,x;m^*)] ,
\end{eqnarray}
and
\begin{eqnarray}
\frac{\delta^2 I}{\delta \pi^a(x) \delta \pi^b(y)} 
\bigg|_{\sigma=\sigma_0,\pi=\pi_0} 
&=& -\frac{1}{2g}\delta^{ab}\delta^4(x-y) \nonumber \\ 
&&+ i\, {\rm tr} [i\gamma_5\tau^a S(x,y;m^*) i\gamma_5\tau^b S(y,x;m^*)] .
\end{eqnarray}

After integrating $\sigma$ and $\pi$
with $\sigma - \sigma_0 \equiv \sigma$ and $\pi -\pi_0 \equiv \pi$
in Eq. (\ref{Z}), we obtain the effective potential at the next to
leading order of the $1/N_c$ expansion, 
\begin{eqnarray}
V(\sigma, \pi) &=& V_0(\sigma, \pi) \nonumber \\
&&+ \frac12 \int \frac{d^4 p}{i(2\pi)^4}
\left\{\ln(\Delta_\sigma^{-1}(p^2;m+\sigma))
+\ln(\Delta_\pi^{-1}(p^2;m+\sigma)) \right\} ,
\end{eqnarray}
here the relation of the minimum of the classical field $\sigma$ and
the classical solution $\sigma_0$ is $\sigma = \sigma_0 + O(1/N_c)$
\cite{Kashiwa:2003rj}.

From the stationary condition and taking $\pi=0$, 
we have the gap equation until next to the leading order,
\begin{eqnarray}
\langle \sigma \rangle 
&=& 2N_f g i{\rm tr}S(m+\langle \sigma \rangle) \nonumber \\
&&- g \int \frac{d^4 p}{i(2\pi)^4} \frac{\partial}{\partial \sigma}
 \left\{\ln(\Delta_\sigma^{-1}(p^2;m+\sigma))
 +\ln(\Delta_\pi^{-1}(p^2;m+\sigma)) \right\}
  \bigg|_{\sigma=\langle \sigma \rangle}.
\label{gap2}
\end{eqnarray}

Substituting the relation of the leading order of the $1/N_c$
expansion for Eq. (\ref{gap2}), we obtain
\begin{eqnarray}
\langle \sigma \rangle &\simeq& 2N_f g i{\rm tr}S(m^*)
 - 2N_f g m^* \int \frac{d^D p}{i(2\pi)^D} 
 \left( \frac{3}{m_{\sigma0}^2 - p^2}+\frac{1}{m_\pi^2 - p^2} \right)\\
&\simeq& \sigma_0 \left( 1 
 - \frac{\delta_\sigma}{N_c} \right) ,
\label{gap3}
\end{eqnarray}
where 
\begin{eqnarray}
\delta_\sigma = \frac{3m_{\sigma0}^2+m_\pi^2}{4m^{*2}} .
\end{eqnarray}
$\sigma_0$ denote the solution of the leading order of the $1/N_c$
expansion. By inserting the values for $m_\pi$, $m^*$ and
$m_{\sigma 0}$, we find $\delta_\sigma \simeq 3.0$.

Similarly, we can obtain the expressions for the chiral condensate and
the sigma mass through a bit of algebra. It may be nice to
summarize the equations in the next to leading order results
which read
\begin{eqnarray}
\langle \bar{u}u \rangle
  &\simeq& \langle \bar{u}u \rangle_0
    \left[ 1 -  \left( 1 - 3 \frac{m}{m^*} \right) 
           \frac{\delta_\sigma}{N_c}            
    \right] , \\
%
 m_{\sigma}^2
 &=& 2(\langle \sigma \rangle + m)^2 
   \left( 2 - \frac{m}{\langle \sigma \rangle} \right) 
 \label{m_sig2} \\
  &\simeq& m_{\sigma 0}^2 - 2 \sigma_0^2
    \left( 4+3\frac{m}{\sigma_0}+\frac{m^3}{\sigma_0^3} \right)
    \frac{\delta_\sigma}{N_c},
 \label{m_sig3}
\end{eqnarray}
these relations are derived from Eqs. (\ref{uu}) and (\ref{pro_sig}).

\subsection{Numerical results}
\label{numeric}

Having obtained the expression for the next to leading order of
the $1/N_c$ expansion, we are now ready for performing the
numerical analysis.

\begin{figure}[!h]
  \begin{center}
    \includegraphics[height=2.0in,keepaspectratio]{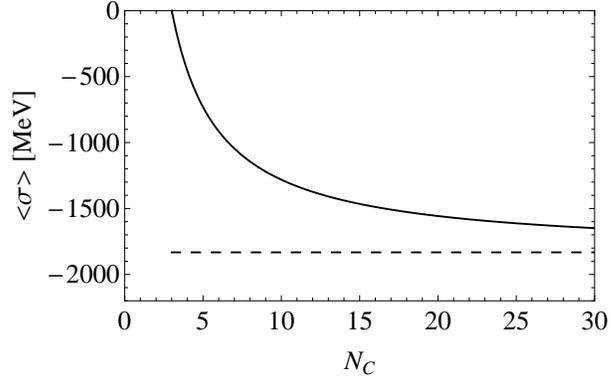} 
  \end{center}
  \vspace{-0.5cm}
  \caption{$N_c$ dependence on $\langle \sigma \rangle$.
  Dashed line: $\sigma_0=-1832.5$MeV.}
\label{fig_sig}
\end{figure}

\begin{figure}[!h]
  \begin{center}
    \includegraphics[height=2.0in,keepaspectratio]{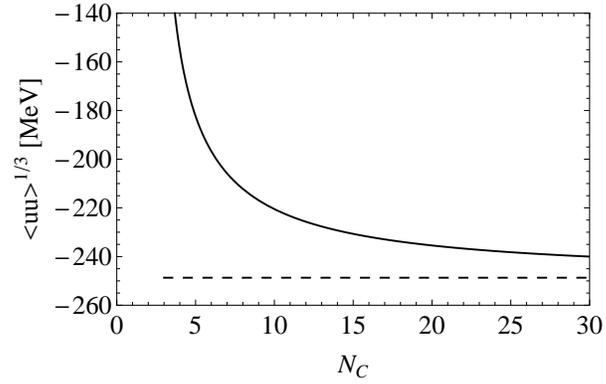} 
  \end{center}
  \vspace{-0.5cm}
  \caption{$N_c$ dependence on $\langle \bar{u}u \rangle^{1/3}$.
  Dashed line: $\langle \bar{u}u \rangle_0^{1/3} =-248.7$MeV.}
\label{fig_uu}
\end{figure}

\begin{figure}[!h]
  \begin{center}
    \includegraphics[height=2.0in,keepaspectratio]{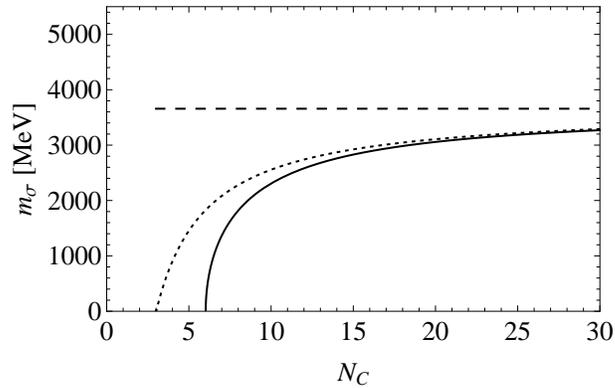} 
  \end{center}
  \vspace{-0.5cm}
  \caption{$N_c$ dependence on $m_\sigma$
    shown in the solid line (from Eq. (\ref{m_sig3})).
    Dashed line: $m_{\sigma} = 3657.5$MeV.
     Dotted curve is the result from the intermediate relation
    in Eq. (\ref{m_sig2}). }
\label{fig_msig}
\end{figure}

Below we show the results for $\langle \sigma \rangle$,
$\langle \bar{u}u \rangle^{1/3}$ and $m_\sigma$ with respect
to $N_c$ in Figs. \ref{fig_sig}, \ref{fig_uu} and \ref{fig_msig}.
The model parameters are determined by the input values:
$m=5.0$MeV, $m_\pi=135$MeV and $f_\pi=92$MeV, as mentioned above.
The solid curves indicate the results for the next to leading order,
and the dashed lines are the ones in the leading order. We see
that $\langle \sigma \rangle$ and $\langle \bar{u}u \rangle$
decrease according to $N_c$, while $m_\sigma$ increases when
$N_c$ becomes large. As trivially expected, the values approach
to the ones in the leading order which are shown in dashed
lines. It is interesting to note that the realistic values
may be found in the region $3<N_c<\infty$, then we think the
expansions performed in the $4$ dimensional limit effectively
work.


\section{Concluding remarks}
\label{conclution}
We have considered the $1/N_c$ correction by taking the four
dimensional limit in the NJL model. We first perform the
calculation up to the next to leading order of the $1/N_c$ expansions.
There we found that the calculations are drastically simplified
thanks to the manipulation of taking the four dimensional limit,
and we are able to study the $N_c$ dependence in the systematic
manner. We also check the numerical tendencies with respect to
$N_c$ on various physical quantities, then show the
predictions of them in which the values approach to that of
the leading order case for large $N_c$.

One important qualitative advantage in this analysis is that
we can perform the next to leading order calculations in an easy
way. While there is quantitative unsatisfactory point on the
value of $m_\sigma$ being considerably larger than the observed
one, as previously found in the corresponding study in which 
the analysis is restricted to the leading order case
\cite{Inagaki:2007dq}. However, the result indicates that the
discrepancy may be cured through including the higher order
corrections since the mass of sigma becomes smaller if we
consider the next to leading order contribution. Therefore, we
think further investigations on the expansion of $N_c$ are
interesting and important.

\section*{Acknowledgments}

HK is supported by the National Research Foundation of Korea
funded by the Korean Government (Grant No. NRF-2011-220-C00011).




\end{document}